\documentclass{raa}            % referee version: for submission

\usepackage{graphicx,times}             %for PS/EPS graphics inclusion, new
\usepackage{natbib}
\usepackage{amssymb,amsmath}
\bibpunct{(}{)}{;}{a}{}{,}

\usepackage[a4paper=true,dvipdfm=true,pagebackref=true]{hyperref}
\hypersetup{colorlinks = true, linkcolor = green, anchorcolor = red, citecolor = blue, filecolor = red, pagecolor = red, urlcolor = red}

\begin{document}

   \title{X-ray Flares Raising upon Magnetar Plateau as an Implication of a Surrounding Disk of Newborn Magnetized Neutron Star
%\footnotetext{$*$ Supported by the National Natural Science Foundation of China.}
}
%   \subtitle{I. Place Your Subtitle Here}

   \volnopage{Vol.0 (20xx) No.0, 000--000}      %%preserved for Editor. DOn't remove!
   \setcounter{page}{1}          %%starting page, preserved for Editor. DOn't remove!

   \author{Tian-Ci Zheng\inst{1}$^{\star}$, Long Li\inst{2}, Le Zou\inst{1}, Xiang-Gao Wang\inst{1}
   %\and B. J. Smith
      %\inst{3}
   }

   \institute{Guangxi Key Laboratory for Relativistic Astrophysics, School of Physical Science and Technology, Guangxi University, Nanning 530004, China; {\it tiancizheng@foxmail.com}\\
        \and
              School of Astronomy and Space Science, Nanjing University, Nanjing 210023, China;\\
        %\and
             %Full institute address for the second author\\
        %\and
             %Full institute address for the third author\\
\vs\no
   {\small Received~~20xx month day; accepted~~20xx~~month day}}

\abstract{
%\del{The X-ray flares identified from near half gamma-ray bursts (GRBs) afterglow. It has known as long-duration, intermittent activities of the central engine}
The X-ray flares have usually been ascribed to long-lasting activities of the central engine of gamma-ray bursts (GRBs), e.g., fallback accretion. The GRB X-ray plateaus, however, favor a millisecond magnetar central engine.
The fallback accretion can be significantly suppressed due to the propeller effect of a magnetar.
Therefore, if the propeller regime cannot resist the mass flow onto the surface of the magnetar efficiently, the X-ray flares raise upon the magnetar plateau would be hinted.
In this work, such peculiar cases are connected to the accretion process of a magnetar, and an implication for magnetar-disc structure is given. We investigate the repeating accretion process with multi-flare GRB\,050730, and give a discussion for the accreting induced variation of the magnetic field in GRB\,111209A. Two or more flares exhibit in the GRB\,050730, GRB\,060607A, and GRB\,140304A; by adopting magnetar mass $M=1.4~ M_\odot$ and radius $R=12~\rm km$, the average mass flow rates of the corresponding surrounding disk are $3.53\times 10^{-4}~M_\odot~\rm s^{-1}$, $4.23\times 10^{-4}~M_\odot~\rm s^{-1}$, and $4.33\times 10^{-4}~M_\odot~\rm s^{-1}$, and the corresponding average sizes of the magnetosphere are $5.01~\rm \times10^{6} cm$, $6.45~\rm \times10^{6} cm$, and $1.09~\rm \times10^{7} cm$, respectively.
A statistic analysis that contains 8 GRBs within 12 flares shows that the total mass loading in single flare is $\sim 2\times 10^{-5}~M_{\odot}$. In the lost mass of a disk, there are about 0.1\% used to feed a collimated jet.
% where less than 0.2\% portion is used to feed a collimated jet.}
}

   \authorrunning{Tian-Ci Zheng}            %author_head in even pages
   \titlerunning{An Implication of Magnetar-disc Structure}  % title_head in odd pages

   \maketitle
%% The author head (on even pages) and the title head (on odd pages) will be
%% automatically extracted from \author{} and \title{}. Whenever the title is too long,
%% you will be asked to supply a shorter one by inserting either \authorrunning{} or
%% \titlerunning{} before \maketitle. Anyway, you can specify your own heads.
%%
%%
%% Note: In the following text body of your manuscript, please note several differences from
%%       other major journals:
%% (1) \subsection{Please Capitalize the First Letter of Each Notional Word in Subsection Title}
%% (2) Please Capitalize the First Letter of Each Notional Word in all tables' captions

%
%________________________________________________ sections below
%
\section{Introduction}           %% first-level sections will be auto-capitalized
\label{sect:intro}
Based on the duration distribution, Gamma-ray bursts (GRBs) are divided into long-duration ($T_{90}>2~\rm s$) and short-duration ($T_{90}<2~\rm s$) types \citep{1993ApJ...413L.101K}. Long GRBs (lGRBs) are believed to originate from the core collapse of a massive star \citep{1993ApJ...405..273W,1998Natur.395..670G,1999ApJ...524..262M,1999Natur.401..453B,2003ApJ...586..356Z} and to associate with the explosion of core-collapse supernovae \citep{2003ApJ...591L..17S,2003Natur.423..847H,2006Natur.442.1008C}. That the merger of two compact objects creates the short GRBs \citep[sGRBs; e.g.,][]{1986ApJ...308L..43P,1989Natur.340..126E,1992ApJ...395L..83N,2005Natur.437..851G,2005Natur.437..859H} was also confirmed by the association of gravitational wave (GW) event, i.e., GW170817 and GRB\,170817A \citep{2017PhRvL.119p1101A,2017ApJ...848L..13A}.
A rapidly rotating and strongly magnetized neutron star (NS) is thought to be born as a central engine, the so called magnetar, whatever lGRBs or sGRBs \citep{1992Natur.357..472U,1994MNRAS.270..480T,1998A&A...333L..87D,2000ApJ...537..810W,2001ApJ...552L..35Z,2010MNRAS.402..705L,2008MNRAS.385.1455M,2011MNRAS.413.2031M,2012MNRAS.419.1537B}. The studies of soft gamma-ray repeater (SGR) showed that the surface dipole magnetic field of a magnetar can be as high as $10^{15}~\rm G$ \citep[e.g.,][]{1998Natur.393..235K,1999ApJ...510L.115K,1999ApJ...519L.139W}. A corotating magnetosphere of the magnetized NS preventing the plasma accretion and throwing away the accreting materials were figured as propeller \citep{1975A&A....39..185I,1998A&ARv...8..279C}.
A competing process of accretion and propeller was figured in the recent studies \citep[e.g.,][]{2005ApJ...623L..41E,2014MNRAS.438..240G,2021ApJ...914L...2L}. However, the coexistence of accretion and outflow was also suggested by some of the magnetohydrodynamic (MHD) simulative studies \citep[e.g.,][]{1997ApJ...489..199G,2005ApJ...635L.165R,2009MNRAS.399.1802R,2018NewA...62...94R,2006ApJ...646..304U}. In these studies, a two-component outflow, propeller-driven conical wind and accretion-feed collimated jet, is seen for a rapidly rotating star scenario.
\cite{2013ApJ...775...67B} attributed the precursor and the prompt emission of GRBs to a jet that forms in the accretion phase.

A canonical GRB X-ray afterglow can be composed of five components \citep{2006ApJ...642..389N,2006ApJ...642..354Z}, in which rapid decay, normal decay and jet break track the properties of a jet, %\del{On the other hand, the X-ray flares are considered as the long-duration activities
however, the shallow decay (or plateau) and the X-ray flares were proposed to connect a long-lasting central engine.
Large sample investigations showed that there are one or more X-ray flares seen in about one third of GRB afterglows \citep{2007ApJ...671.1921F,2007ApJ...671.1903C,2010MNRAS.406.2113C}, the time domain analyses for its lightcurve imply that the X-ray flares have an ``internal" origin, and a new idea for restart the central engine is required \citep{2005Sci...309.1833B,2005MNRAS.364L..42F,2006ApJ...646..351L,2007MNRAS.375L..46L}.
An intermittent hyperaccreting disk surrounding a black hole (BH) seems to be a great candidate \citep{2005ApJ...630L.113K,2006ApJ...636L..29P}. \cite{2014ApJ...789..129C} imported a competition between the magnetic field and the neutrino dominated accretion flow (NDAF) to avoid a fragmentary hyperaccreting disk. By comparing with the diffusion timescale of magnetic field, a time interval between the two successive accreting episodes is about 0.2 s, and the mass accretion rate can be as high as $\sim 1~M_\odot\rm~s^{-1}$ for a external magnetic field $B=10^{14} ~\rm G$.
Interestingly, the initially accumulated shells onto blast wave to produce the observed shallow decay was also investigated. However, the observed X-ray afterglows are too dark as compared to the prediction of the model \citep{2009ApJ...707.1623M}.
A statistic analysis of {\em Swift}/XRT data showed that the lightcurve decay slope $-0.75$ distinguishes between the normal decay segment and the shallow decay segment \citep{2008ApJ...675..528L}.
A continuous energy injection into the forward shock explains the shallow decay successfully \citep{2009MNRAS.397.1177E}.
However, the X-ray plateau followed by sharp decay in some cases invoke an explanation with the internal energy dissipation of a magnetar wind \citep[e.g.,][]{1990ApJ...349..538C,1992Natur.357..472U,1994MNRAS.267.1035U,2007ApJ...665..599T,2010MNRAS.409..531R,2013MNRAS.430.1061R,2014MNRAS.443.1779R,2014ApJ...785...74L,2015ApJ...805...89L}. The end of the magnetar plateau is featured as a steeper decay for the spindown process of a magnetar or a very steep decay for the collapse of magnetar into a BH,
where the typical slopes are $\sim-2$ \citep{2001ApJ...552L..35Z} and $<-3$ \citep{2007ApJ...670..565L,2010MNRAS.402..705L,2010MNRAS.409..531R}, respectively. Then, a re-brightening catching the end of the very steep decay symbolizing the fallback material onto the newborn BH was proposed \citep{2017ApJ...849..119C}.

The connection between X-ray flares and the millisecond magnetar has suggested by \cite{2006Sci...311.1127D}, but more credible  evidence is expected to be excavated. Some peculiar cases display a coexistence of X-ray flares and magnetar plateau, which persuade us to connect between the magnetar accreting from the surrounding disc and the spindown process of a magnetar.
If the magnetic dipole (MD) radiation of a magnetar is in progress,
the charged particles would form a magnetosphere surrounding the magnetar due to the affection of strong magnetic field.
The propeller regime keeps away the fallback material to form a dense disk next to the magnetosphere. However, an unstable channelled flow would be led by gravitational force along the magnetic field line onto the polar cap of the central magnetar.
In the meantime, owing to the magnetic and centrifugal forces, a small fraction of the accreting material penetrates into the opened polar magnetic field line and feeds a relativistic collimated jet, then a considerable X-ray flare is induced.
In this work, we argue that the X-ray flares raising upon the magnetar plateau can be used to connect the accretion process of a magnetar and to lead an implication for magnetar-disc structure. The properties of the jet, magnetar, and disk are investigated with both the special cases and a small sample study. In Section 2, we review the activities of newborn magnetized NS, both case study and sample analysis are presented in Section 3,
conclusions and discussions are organized in Section 4. A $\Lambda$CDM cosmology with parameters $H_0 = 70 ~\rm {km\ s^{-1}\ Mpc^{-1} }$, $\Omega_M= 0.30$, and $\Omega_\Lambda = 0.70$ is adopted.

%% Authors can give a citation as 'Michel et al. 1992'.
%% You may also use \cite, \citep and \citet for citation, and use Table~1 or Figure~1
%% and so forth. Using \ref and \label for cross-references of Tables/Figures
%% is a good way in adjusting/adding/removing text, tables or figures.

\section{Millisecond Magnetar Activities}
\label{sect:MMA}
By adopting a rotating progenitor model \citep{2000ApJ...528..368H}, the particle hydrodynamics simulation showed that the initial period $P_0$ of a newly born NS is $\sim 100$ ms, and drops to $\sim2$ ms after a short cooling  \citep{2000ApJ...541.1033F}.
A considerable rotational energy is stored in newly born millisecond NS,
\begin{eqnarray}%equnarry
E_{rot}=\frac{1}{2}I\Omega^2 \simeq 2\times 10^{52}M_{1.4}R_6^2P_{-3}^{-2}~\rm erg,
\label{eq:E_rot}
\end{eqnarray}
where $I$ is the moment of inertia, for a NS holds mass $M$ and radius $R$, it can be written as $I = 0.35MR^2$ \citep{2001ApJ...550..426L}. The spin period $P=2\pi/\Omega$, where $\Omega$ is the angular frequency of the NS, and the notation $Q_n = Q/10^n$ in cgs units, furthermore $M_{1.4}=M/1.4 M_{\odot}$. For a magnetar central engine scenario,
its rotational energy lose as a magnetar wind or an ejecta and is observed in the X-ray afterglows,

\begin{eqnarray}%equnarry
-\dot{E}_{rot}=I\Omega\dot{\Omega}=L=f_b L_{\rm iso,X}/\eta,
\label{eq:Edot}
\end{eqnarray}
where $L_{\rm iso,X}$ is isotropic X-ray luminosity. The efficiency of rotational energy to the observed X-ray emission $\eta$ holds significantly different value for different process. For the {\em Swift}/XRT \citep{2005SSRv..120..165B}, about $1\%$ jet energy can emit into the observation window in a GRB prompt emission analogous to the GRB radiative efficiency \citep[e.g.,][]{2007ApJ...655..989Z,2015ApJS..219....9W},
and as high as $50\%$ rotational energy can be observed for a magnetar wind \citep{2011MNRAS.413.2031M,2014ApJ...785...74L}.
$f_b=1-$cos$\theta$ is the beaming factor, where $\theta$ is opening angle. The opening angle of a magnetar wind $\theta_{\rm dip}$ is also worth to budget. It should larger than a jet of GRBs $\sim0.1~\rm rad $ \citep[$\sim 6^{\circ}$;][]{2001ApJ...562L..55F}, but smaller than a low speed conical wind $\sim 30^{\circ}-40^{\circ}$ \citep{2009MNRAS.399.1802R}.
%\del{A suitable study result of GRB magnetar central engines shows that beaming angle is about $\theta_{dip} \sim 15^{\circ}~(f_b=0.034)$ by adopting the observed plateau luminosity holds $\sim20\%$ ($\eta_{dip}=0.2$) rotational energy (Rowlinson et al. 2014). A more comprehensive study of the conversional efficiency shows that the observed X-ray afterglow portion typical less than $10\%$ (Xiao \& Dai 2019), but we do not adopted it and the reason is given in Section \ref{sect:CaD}.}
A study for the internal dissipation of magnetar wind showed that the observed X-ray emission typical less than $10\%$ when a  saturation Lorentz $\Gamma_{sat} \geq 100$ is adopted \citep{2019ApJ...878...62X}. According to the study of GRB magnetar central engines by \cite{2014MNRAS.443.1779R}, when adopting the observed plateau holds $\sim 5\%$ ($\eta_{\rm dip}=0.05$) rotational energy, the beaming angle is $\sim 10^{\circ}$ ($f_b=0.015$).
Furthermore, the observed X-ray energy fraction $\eta_{\rm fla}=0.05$, and a moderate beaming angle $\theta_{\rm fla}=10^{\circ}$ for the X-flares are adopted in this work.

\subsection{Magnetic Dipole Radiation}
For a magnetar holds surface magnetic field strength $B$ and initial spin period $P_0$, its spindown luminosity $L$ is featured as a plateau ($t\ll\tau$) followed by a sharp decay ($t\gg\tau$) \citep{2001ApJ...552L..35Z},
\begin{eqnarray}
L=L_0(1+\frac{t}{\tau})^{-2},
\label{eq:observation}
\end{eqnarray}
where the characteristic spindown luminosity $L_0$ and the timescale $\tau$ is given by
\begin{eqnarray}
L_{0}&=&1.0\times 10^{49}B_{15}^2P_{0,-3}^{-4}R_6^6 ~\rm erg~s^{-1},\nonumber\\
\tau&=&2.05\times10^{3}I_{45}B_{15}^{-2}P_{0,-3}^2R_6^{-6} ~\rm s.
\label{eq:spin_down}
\end{eqnarray}
When considering the affection of multiple energy dissipating mechanisms and the spectral evolution of radiative process, the slope of a sharp decay may hold a very different value \citep{2017ApJ...843L...1L,2018MNRAS.480.4402L,2019ApJ...871...54L,2019ApJ...878...62X}.

\subsection{Accretion Process of a Magnetar}
The surface magnetic field strength of millisecond magnetar is as high as $\sim10^{15}$ G. It would affect the ionized materials that close to the surface of magnetar and boost the forming of a corotating magnetosphere. The material inside the magnetosphere is dominated by the magnetic pressure, and the magnetic pressure in any given radius $r$ is written as $P_m=\mu^2/8\pi r^6$, where the MD moment of the magnetar is written as $\mu=BR^3$.
An accretion flow from the disk exerts a ram pressure $P_{\rm ram}=\dot{M}_{\rm disk}(2GM)^{1/2}/{8\pi r^{5/2}}$, where $\dot{M}_{\rm disk}$ is the mass flow rate of the disk, and $G$ is the gravitational constant. Therefore, the position where the material pressure comparable with the magnetic pressure is defined as Alv{\'e}n radius \citep{1973ApJ...179..585D,2005ApJ...623L..41E},
\begin{eqnarray}
R_A&=&\left(\frac{\mu^4}{2GM\dot{M}_{\rm disk}^2}\right)^{1/7}\nonumber\\
&=&2.5\times 10^6M_{1.4}^{-1/7}R_6^{12/7}B_{14}^{4/7}\dot{M}_{\rm disk,-5}^{-2/7}~\rm cm.
\label{eq:Alfven_radius}
\end{eqnarray}
At the corotating radius $R_c$, the material keeps the same angular frequency with magnetar without magnetic force considered,
\begin{eqnarray}
R_c&=&\left(\frac{GM}{\Omega^2}\right)^{1/3}\nonumber\\
&=&1.7\times10^6M_{1.4}^{1/3}P_{-3}^{2/3}~\rm cm.
\label{eq:corotating_radius}
\end{eqnarray}

Hence, when the magnetic pressure can not keep the disk as far as corotating position, e.g., $R_A< R_c$, an unstable channelled flow would be led by gravitational force along the magnetic line onto the polar region of the central magnetar.
There are two parts of the energy transmitted from accretion materials to a magnetar, the kinetic energy and the gravitational potential energy. Considering a continuous constant accretion rate, the energy transmission can be written as
\begin{eqnarray}
\dot{E}_{acc}&=&E_{\rm start}-E_{\rm end} \nonumber\\
&=&\frac{1}{2}\dot{M}_{\rm acc}\left(R_A^2\Omega_{\rm K,A}^2-R'^2\Omega^2\right)\nonumber\\
&~&-GM\dot{M}_{\rm acc}\left(\frac{1}{R_A}-\frac{1}{R}\right),
\label{eq:acc_energy_1}
\end{eqnarray}
where $\dot{M}_{\rm acc}$ is mass flow rate of the accretion onto the magnetar, $\Omega_{\rm K,A}=(GM/R_A^3)^{1/2}$ is the Keplerian frequency at the Alfv\'en radius, and one has $\Omega_{\rm K,A}>\Omega$ for $R_A< R_c$. When the accretion materials inhabit the polar cap of the magnetar, the radius of its moment of inertia is marked as $R'$, and here has $R'\leq R<R_A$. For a typical NS with surface magnetic field $B=10^{15}~\rm G$ and a mass flow rate of the disk $\dot{M}_{\rm disk}<10^{-3}~M_{\odot}$, one has $R_A>2.5R$.
Therefore, the approximate result is given as
\begin{eqnarray}
%E_{acc}&\approx&\frac{1}{2}\dot{M}_{acc}R_A^2\Omega_{K,A}^2+GM\dot{M}_{acc}\left(\frac{1}{R}-\frac{1}{R_A}\right)\nonumber\\
\dot{E}_{\rm acc}&\approx&\frac{1}{2}GM\dot{M}_{\rm acc}\frac{1}{R_A}+GM\dot{M}_{\rm acc}\left(\frac{1}{R}-\frac{1}{R_A}\right)\nonumber\\
&\approx&GM\dot{M}_{\rm acc}\frac{1}{R}.
\label{eq:acc_energy_2}
\end{eqnarray}
Hence, we find that the most of energy is donated by gravitational potential energy.
Accreting materials transmit its kinetic and gravitational potential energy onto the magnetar, then the latter spins up.

A magnetized, rapidly rotating star accompanied by an accretion disk was studied by the MHD simulative investigations \citep[e.g.,][]{1997ApJ...489..199G,2005ApJ...635L.165R,2009MNRAS.399.1802R,2018NewA...62...94R,2006ApJ...646..304U}. These studies suggested that the mass flow to the star and to the outflow can take place at the same time, and a two-component outflow, propeller-driven conical wind and accretion-feed collimated jet, can be seen in the polar region. In the accretion phase, most of funnelled accreting flow on to the pole of star.
However, owing to the magnetic and centrifugal force, a part of it flows into the opened magnetic lines and feeds a collimated jet \citep{1999ApJ...524..159G,2009MNRAS.399.1802R}.
% a part of it, however, flow into the opened magnetic lines due to the magnetic and centrifugal force.
The mass flow rates rely on the magnetar parameters, and the relations are approximated as \citep{2006ApJ...646..304U}
\begin{eqnarray}
&\dot{M}_{\rm wind}&\propto\Omega^{2.6}\mu^{0.9},\nonumber\\
&\dot{M}_{\rm jet}&\propto\Omega\mu^{0.2},\nonumber\\
&\dot{M}_{\rm acc}&\propto\Omega^{-5}\mu^{-1.3},\nonumber\\
&\dot{M}_{\rm disk}&\propto\Omega^{-2.2}\mu^{0.9},
\label{eq:mass_trace}
\end{eqnarray}
where, the mass flow rates of the collimated jet and the conical wind are marked as $\dot{M}_{\rm jet}$ and $\dot{M}_{\rm wind}$, respectively.
%This relation is based on the simulation result of a relevant steady star-disc structure and an extremely low value of MD moment $5\leq\mu\leq20$ comparing with a magnetar \add{The describe has some problem !}.
%Further more, the mass flow rate of accretion onto the magnetar $\dot{M}_{\rm acc}$ with a fast raising and exponential decaying like profile in single accretion event maybe more suitable (MacFadyen et al. 2001; Zhang et al. 2008; Dai \& Liu 2012).
Comparing with the propeller-driven conical wind, the low-density, high-velocity, magnetic dominated, collimated jet is more energetic.
%Here we more focus on the mass traces.
For a protostar, if considering the jet is a component that poloidal velocities $\nu_p > 1.5 \nu_K$, where $\nu_K$ is Keplerian velocity, the jet component carries about $1\%$ of the total outflowing mass and $13\%$ of the total lost angular momentum of the star \citep{2009MNRAS.399.1802R}. Therefore, it is easy to get that about $69\%$ lost rotational energy is carried by the jet component. A same outflow scenario is adopted in this work.

In the propeller regime, a propeller efficiency is organized as \citep{2018NewA...62...94R}
\begin{eqnarray}
f_{\rm eff}=\frac{\dot{\bar{M}}_{\rm out}}{\dot{\bar{M}}_{\rm acc}+\dot{\bar{M}}_{\rm out}}
\label{eq:pro_effiency}
\end{eqnarray}
where $\dot{\bar{M}}_{\rm out}$ and $\dot{\bar{M}}_{\rm acc}$ are time-averaged mass flow rate of the outflow and accretion onto the star.
$\dot{\bar{M}}_{\rm out}$ relates to a considered minimum outflowing poloidal velocities $\nu_{\rm min}$,
when $\nu_{\rm min} > \nu_{\rm esc}$ is considered, the average propeller efficiency has $f_{\rm eff}=0.0006\omega^{4.01}_s$, where $\nu_{\rm esc}$ is escape velocity and $\nu_{\rm esc}\sim 1.5 \nu_K$, $\omega_s=\Omega/\Omega_{K,A}$ is a fitness parameter \citep{2018NewA...62...94R}.
In this scenario, the outflow can ascribe to the collimated jet, that is $\dot{M}_{\rm out}=\dot{M}_{\rm jet}$, where $\dot{M}_{\rm out}$ is mass flow rate of the outflow.
By adopting $\omega_s=1.2$, the accretion mass flow rate and the mass lost rate of disk can be estimated with the mass flow rate of a jet,
%the connection that the accretion mass flow rate with the mass flow rate of jet and the mass lost rate of the disk with the mass flow rate of jet the roughly estimate with
\begin{eqnarray}
\dot{M}_{\rm acc}=832\dot{M}_{\rm jet},\nonumber\\
\dot{M}_{\rm dis}=931\dot{M}_{\rm jet}.
\label{eq:jet_acc}
\end{eqnarray}
The one should note that the disk oscillations were also suggested by the mentioned MHD simulative studies, therefore the $\omega_s$ may get a prominent change at the before and after the accretion process.

Since some studies have suggested that the central magnetar could spin up by the accretion toque and spin down by the propeller torque \citep[e.g.,][]{2005ApJ...623L..41E,2014MNRAS.438..240G,2021ApJ...914L...2L}, it is necessary to budget the energy income and output of a magnetar.
The total mass loading in the jet $M_{\rm jet}=\int{\dot{M}_{\rm jet}}dt$ can be estimated with $E_{\rm fla}=M_{\rm jet}c^2\gamma$, where $c$ is the speed of the light, $E_{\rm fla}$ is the total energy for a single event, and $\gamma$ is the bulk Lorentz factor of the jet. By adopting the bulk Lorentz factor of the jet component $\gamma=100$ \citep{2009ApJ...707.1623M}, the lost and gathered energy of the magnetar can be budgeted, $\zeta=(E_{\rm jet}/0.69)/E_{\rm acc}\sim 3$. Since too many uncertainty in the estimation, and a not significant advantage is charged by the lost energy, the spin evolution of the central magnetar is not considered in this work.
To estimate mass loading in the jet, the $E_{\rm fla}$ can be derived with the observed isotropic luminosity of X-ray flares $L_{\rm fla,X,iso}$, that is
\begin{eqnarray}
E_{\rm fla}= f_b \int{L_{\rm fla,X,iso}}dt/\eta_{\rm fla}.
\label{eq:jet_mass}
\end{eqnarray}
\rm

%\section{Sample study: GRB\,060607A, GRB\,050730}
\section{Sample study and the tests of physical parameters}
The X-ray afterglow data are derived from the UK {\em Swift} Science Data centra \citep[UKSSDC;][]{2009MNRAS.397.1177E}\footnote{http://www.swift.ac.uk/results.shtml}. The selected cases present a X-ray plateau followed by a sharp decay, which is identified as the internal energy dissipation of a magnetar wind. Ahead of the sharp decay, the prominent X-ray flares, the internal shock \citep{2009ApJ...707.1623M} which originates from the accreting fed collimated jet, raise upon the magnetar plateau.
%\del{The selected sample presents a prominent sharp decay ($\alpha<-2$) following a long-duration plateau ($\alpha<0.75$), which is attributed to the internal dissipation of MD wind. Ahead of the sharp decay, remarkable X-ray flares, the internal shock (Maxham \& Zhang 2009) from the accreting fed magnetic jet, raise upon the MD emission.}\del{The sample is organized in Table \ref{Tab:Magnetar} and shown in Figure \ref{fig:samples}.}
The lightcurves are fitted with a smooth broken power law function
\begin{eqnarray}
F=F_0 \left[\left(\frac{t}{t_b}\right)^{-\omega\alpha_1}+\left(\frac{t}{t_b}\right)^{-\omega\alpha_2}\right]^{-1/\omega},
\label{eq:BPL}
\end{eqnarray}
where the sharpness parameter marked as $\omega$, the constant flux at break time $t_b$ is written as $F_b=F_0\cdot 2^{-1/\omega}$, and the decay indices before and after $t_b$ is described as $\alpha_1$ and $\alpha_2$.

\subsection{GRB\,050730}
\label{sect:case-050730}
\begin{figure}[h]
\centering
\includegraphics[angle=0,scale=0.350]{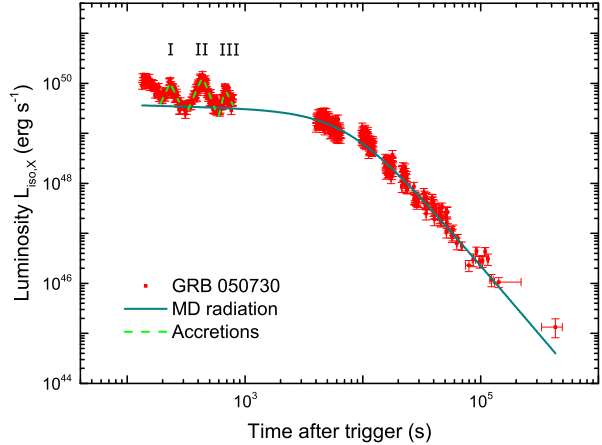}
\caption{Lightcurve of GRB\,050730, the multi-flare phenomenon is carried to connect the repeating accretion process of a magnetar.}
\label{fig:luminosity-050730}
\end{figure}

The weak lGRB\,050730 triggered {\em Swift}/BAT at 19:58:23 on 2005-07-30 \citep{2005GCN..3704....1H} ($T_0$ in the following), with prompt emission duration $T_{90}(15-150~\rm keV)=157\pm 18~s$ at redshift $z=3.967 $ \citep{2005GCN..3709....1C,2005GCN..3710....1R,2005GCN..3716....1H,2005GCN..3732....1P,2005GCN..3746....1D}. The average photon index for its mean photon arrival time $T_0+17681$ $\rm s$ is $\Gamma=1.58$. Its X-ray afterglow featured as three X-ray flares competing with the long-duration plateau and followed by a sharp decay ($\alpha_2=2.77$).
We derive the isotropic luminosity by using $L(t)=4\pi D_L^2f(t)k(z)$ where $D_L$ is the luminosity distance, $f(t)$ is the observed X-ray flux and $k(z)=(1+z)^{\Gamma-2}$ is the cosmological $k$-correction factor \citep{2001AJ....121.2879B,2019ApJ...886....5S}. The lightcurve is presented in Figure \ref{fig:luminosity-050730}.
For the MD radiation model, the characteristic spindown luminosity and time scale are $L_0 = (2.81\pm0.49)\times 10^{49}\times f_b/\eta  ~\rm erg~\rm s^{-1} $ and $\tau=(7.65\pm0.49)\times 10^3/(1+z)~\rm s$, respectively. The surface magnetic field and the initial spin period can be estimated with Eq. (\ref{eq:spin_down}), and ones have $B=(1.16\pm0.13)\times 10^{15}~\rm G$ and $P_0 = (1.48\pm0.14)~\rm ms$. The observed X-ray fluence of the flares integrate since $T_0+193~\rm s$, $T_0+327~\rm s$, and $T_0+598~\rm s$ for flare $\rm\uppercase\expandafter{\romannumeral1}$, flare $\rm\uppercase\expandafter{\romannumeral2}$ and flare $\rm\uppercase\expandafter{\romannumeral3}$, and ones have $7.24\times 10^{-8} ~\rm erg~\rm cm^{-2}$, $2.08\times 10^{-7} ~\rm erg~\rm cm^{-2}$, and $1.23\times 10^{-7} ~\rm erg~\rm cm^{-2}$, respectively.

The jet forming process is still ambiguous. In Section \ref{sect:MMA}, we assume that the accretion onto the magnetar feeds a collimated jet to create the considerable X-ray flares. In this scenario, the accreting material supplies its kinetic energy and gravitational potential energy to keep the period of the central magnetar without a significant change, the total energy of the observed X-ray flare equal to those gravitational potential energy quantitatively.
By adopting a magnetar with mass $M=1.4~M_\odot$ and radius $R=12~\rm km$, the total mass loading in each flare is estimated. The mass loading in flares $\rm\uppercase\expandafter{\romannumeral1}$, $\rm\uppercase\expandafter{\romannumeral2}$, and $\rm\uppercase\expandafter{\romannumeral3}$ are $9.28\times 10^{-6}~ M_\odot$, $2.67\times 10^{-5}~ M_\odot$, and $1.57\times 10^{-5}~ M_\odot$, respectively.

The collected peak time $t_p$ (time after BAT trigger) of the three successive flare $\rm\uppercase\expandafter{\romannumeral1}$, $\rm\uppercase\expandafter{\romannumeral2}$, and $\rm\uppercase\expandafter{\romannumeral3}$ are 230 $\rm s$, 435 $\rm s$, and 677 $\rm s$, respectively.
The time interval of the first flare to the trigger time almost equal to the time interval of the adjacent two X-ray flares quantitatively. In the rest frame, they are 46 $\rm s$, 41 $\rm s$, and 49 $\rm s$ respectively.
Owing to the MHD instability of the disk \citep[e.g., Rayleigh-Taylor, Kelvin-Helmholtz, and Balbus-Hawley instabilities; see][]{1991ApJ...376..214B,1992ApJ...400..610B}, the disk would get prominent oscillations, the inner disk could be stripped and be accreted onto the magnetar finally \citep[e.g.,][]{1997ApJ...489..890M,1999ApJ...524..159G,2002ApJ...578..420R}.
Here, we connect the X-ray flares to the periodic accretion process; it is a four-step cycle starting from accretion to quiescent then restart again \citep{1999ApJ...524..159G,2018NewA...62...94R}.
Firstly, the materials accrete at the inner disk and then move inward gradually, the magnetosphere is compressed by the accreted materials in this phase. Secondly, the material at the inner disk penetrates across the outer region of the magnetosphere. During this phase, the inflation of magnetic field line occurs, it leads the reconnection of field lines. Then a part of the material flows to the surface of magnetar, and a apart of it is ejected. Finally, the magnetosphere is cleaned, then it expands and travels ahead of the steps once more.
%First, the materials are collected and to help the accreting of the disk, the material at the inner disk start to move inward and penetrates into the external regions of the magnetosphere, at this phase, the magnetic field line start inflation. Then a part of materials goes onto the surface of NS and partly as the outflow ejected. Finally, the magnetosphere is cleaned, than begin to expend and travels ahead the steep once more.
In this scenario, the quiescent period $\Delta t$ in a high diffusivity scenario can be limit with \citep{2018NewA...62...94R}
\begin{eqnarray}
\Delta t > \frac{\mu^2\Delta r}{\dot{M}_{\rm disk} R^3_A G M(\Omega/\Omega_d-1)^2},
\label{eq:time_interval}
\end{eqnarray}
where $\Delta r$ is the depth that material penetrates into the magnetosphere, and $\Omega_d$ is the angular frequency of a disk. $\Delta r $ may comparable with the size of magnetosphere, since a significant variation of magnetosphere is shown in the simulation \citep{2018NewA...62...94R}. By adopting $B=10^{15} ~\rm G$, $ \dot{M}_{\rm disk}=10^{-4}~ M_{\odot}~\rm s^{-1}$, $M=1.4 ~ M_\odot$, $R=1.2\times 10^6 ~\rm cm$, $R_A=2\Delta r =6\times 10^6~\rm cm$, and $\Omega/\Omega_d - 1 \sim 0.2$, the limit has $\Delta t >0.03~\rm s$.
In the low diffusivity scenario, the time interval exhibits a linear dependence with diffusivity coefficient, and the power of other coefficients twice as high as those in the high diffusivity scenario is required \citep{2018NewA...62...94R}. The unsteady collimated jets yield in lower diffusive flows was suggested by \cite{1999ApJ...524..159G}.
%Furthermore, to consider the time interval, the processes that the matter accumulates and move inward may also important (see Figure (B.18) in Romanova et al. 2018), these processes should deep rely on the disk mass flow rate and the state of central magnetar, i.e., spin and surface magnetic field.}

%\add{The periodic accretion process can also implicant that the disk has reached a quasi-steady state, the MHD instability leading the oscillation of disk, and trigger a short-duration accretion. Just as the simulation in Li et al. 2020, after a significant accretion, the ratio of propeller toque to accretion keeps close to 1, without the absolutely advantage of accretion or propeller, propeller and accretion reach dynamic equilibrium.}

%\add{Following the time interval of the two episodes of the adjacent flares, The average mass lost rate of the disk for GRB\,050730, is $3.44\times 10^{-5}~\rm M_\odot ~s^{-1} $. Therefore, by using the Eq. (\ref{eq:Alfven_radius}), the size of magnetosphere is given as $20.5R$.}

\subsection{GRB\,111209A}
\label{sect:case-111209A}
The supernova associated lGRB\,111209A at redshift $z=0.677$ \citep{2011GCN.12648....1V}, an Ultra-long duration of the prompt emission $T_{90} \sim 25000~\rm s$ was suggested by \cite{2013ApJ...766...30G}. The burst should explode at least $2000 ~\rm s$ ahead of the BAT trigger, and a significant precursor start at $\sim T_0-5000 ~\rm s$ \citep{2011GCN.12663....1G}.
Its X-ray observations exhibit a long-duration plateau followed by a sharp decay ($\alpha=4.50$), but a prominent X-ray flare exhibits at $T_0 + 2000~\rm s$. After suffering a dark ages as long as 3000 $\rm s$, the luminosity from dark climb to normal level accompanying with a small fluctuation.
The characteristic spindown luminosity and time scale are $L_0 = (3.77\pm0.07)\times 10^{48}\times f_b/\eta ~\rm erg~s^{-1} $ and $\tau = 1.23\pm0.03\times 10^4/(1+z) ~\rm s$, respectively.
The surface magnetic field and the initial spin period are $B = 0.67\pm 0.02\times 10^{15}~\rm G$ and $P_0 = 1.85\pm0.03~\rm ms$. Due to an inadequate observation, a putative profile of the observed X-ray flare is adopted to estimate the mass loading, the given mass loading is $1.72\times 10^{-5}~ M_\odot$.
%\del{The significant luminosity from dark climb to normal level at the MD dipole radiation duration, hinting the process of MD dipole radiation is suppressed.}
\begin{figure}[ht]
\centering
\includegraphics[angle=0,scale=0.350]{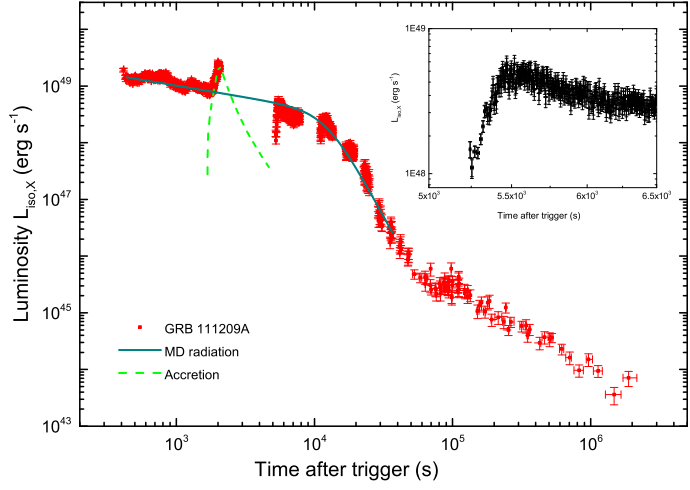}
\caption{The lightcurve of GRB\,111209A. A putative fast raising and exponential decaying profile is used to make a demonstration of accretion profile, and the inset presents a prominent luminosity from dark climb to normal level.}
\label{fig:luminosity}
\end{figure}

In principle, as the accretion materials close to the surface of the NS, the magnetic field would be dragged by the accretion materials, but the diffusion of magnetic field is also ongoing. If an absolute advantage is occupied by the accretion materials, the magnetic field lines could be buried. Hence, the surface magnetic field of NS would experience a significant decrease, and an empirical behaviour can be written as \citep{1986ApJ...305..235T,1989Natur.342..656S,2013ApJ...775..124F}
\begin{eqnarray}
B(t)=\frac{B_{i}}{1+M_{\rm acc}/M_c},
\label{eq:magnetic_bury}
\end{eqnarray}
where $B_{i}$ is the initial surface magnetic field strength of NS, $M_{\rm acc}$ the total material that accretes onto magnetar, and the critical mass $M_c$ ranges from $10^{-5}$ to $10^{-3}~ M_{\odot}$.
%\add{According to the observed X-flare at $T_0+2000$, We have the total mass accreting onto the surface of magnetar is $6.60\times 10^{-3} ~\rm M_{\odot}$. }
By taking $M_c=10^{-3}~ M_{\odot}$, the surface magnetic field of GRB\,111209A should decrease from $0.67\times 10^{15} ~\rm G$ to $0.04\times 10^{15}~\rm G$, and the observed magnetar plateau would experience a significant shrink, for which $L_{\rm dip}\propto B^2$.
The significant luminosity from dark climb to normal level of the magnetar plateau was observed at $\sim T_0+5000~\rm s$ and $\sim T_0+10000 ~\rm s$,  and an inset in Figure \ref{fig:luminosity} gives a demonstration. Whether it can be connected with a re-magnetized process of NS is under debate.
The buried magnetic field is suppressed, as time goes by, the NS should be re-magnetized again \citep{1999A&A...345..847G}, and the MD radiation would go back to the normal level.
The re-magnetized process that based on Ohmic diffusion and the Hall drift should experience thousands years or more \citep{1999A&A...345..847G,2011MNRAS.414.2567H,2013ApJ...775..124F}. However, since the magnetar just born thousands of even hundreds of seconds, one of the process that amplifies the initiate magnetic field of a magnetar, convection in the magnetar, cannot be neglected. As estimated by \cite{1993ApJ...408..194T}, this process would take about $30 ~\rm s$ only. Hence, it may dominate the re-magnetized process of a buried magnetar scenario.
On the other hand, owing to the affection of magnetic field, the funnelled accreting flow is located in the polar region of the magnetized NS \citep{1973ApJ...184..271L,1977ApJ...215..897E,2002ApJ...578..420R}, the accreting materials onto the centre of the polar cap would also be resisted by the opened magnetic field line \citep{2009MNRAS.399.1802R}. Therefore, the re-magnetized process of a portion-buried NS would proceed at the both interior and exterior, and the surface magnetic field should be less affected by interior one.
In this scenario, those observed phenomena that the ligtcurve from dark climbs to normal in the magnetar plateau is expected to be understood as the re-magnetized process of a magnetar.

Of course, a significant emission at the period $T_0+2000$ to $ T_0+ 5000$ $\rm s$ was detected by Konus-Wind \citep{2011GCN.12663....1G}, and a significant transition from bright to dark is also presented. If the suppressing process of magnetic field corresponds to the flare at $T_0+2000$, then the emission may be explained as the afterglow component, which is always covered by the magnetar plateau in the observed band of {\em Swift}/XRT \citep[e.g., GRB\,070110;][]{2007ApJ...665..599T}. If the suppressed magnetic field is caused by the accretion at this phase, it may give a hint that a re-magnetized process can be as short as hundreds of seconds.

%Whether the this part of the emission is from the afterglow still under debate, for what also covered by the magnetar plateau in the observed band of {\em Swift}/XRT detector (e.g., GRB\,070110; Troja et al. 2007), the buried magnetic line is caused by this part of the accretion also can not be excluded.}

\begin{table}
\renewcommand{\thetable}{\arabic{table}}
\centering
\caption{MD radiation.}
\centering
\label{Tab:Magnetar}
\footnotesize
\begin{tabular}{lccccccc}
%\tablewidth{0pt}
\hline\hline
GRB	&	redshift $z$	&	$\Gamma$	&	$-\alpha_2$	&	$\tau (z)~(\rm \times 10^{4} ~s)$		&	$L_{\rm 0,X,iso}~(~\rm erg~s^{-1})$	&	$B~(\rm \times 10^{15} ~\rm G)$	&	$P_0~(\rm ms)	$			\\	
050730	&	3.967$^{(1)}$ 	&	1.58	&	2.77 	&	0.77	$\pm$	0.05	&	(2.81	$\pm$	0.49)$\times 10^{49}$	&	1.16 	$\pm$	0.13 	&	1.48 	$\pm$	0.14 \\
060607A	&	3.082$^{(2)}$ 	&	1.55	&	3.45 	&	1.27	$\pm$	0.03	&	(2.35	$\pm$	0.11)$\times 10^{48}$	&	1.98 	$\pm$	0.07 	&	3.58 	$\pm$	0.09 \\
111209A	&	0.677$^{(3)}$ 	&	1.79	&	4.50 	&	1.23	$\pm$	0.03	&	(3.77	$\pm$	0.07)$\times 10^{48}$	&	0.67 	$\pm$	0.02 	&	1.85 	$\pm$	0.03 \\
140304A	&	5.283$^{(4)}$  	&	1.96	&	3.50	&	0.18	$\pm$	0.04	&	(4.53	$\pm$	0.88)$\times 10^{49}$	&	4.98 	$\pm$	1.19 	&	2.71 	$\pm$	0.40 \\
201221A	&	5.7$^{(5)}$  	&	1.45	&	2.70 	&	1.44	$\pm$	0.58	&	(8.94	$\pm$	0.82)$\times 10^{47}$	&	4.66 	$\pm$	1.89 	&	7.00 	$\pm$	1.45 \\
\hline
060413	&	...	&	1.53	&	3.06 	&	2.51	$\pm$	0.13	&	(1.49	$\pm$	0.11)$\times 10^{48}$	&	1.47 	$\pm$	0.09 	&	3.45 	$\pm$	0.15 \\
071118	&	...	&	1.59	&	2.20 	&	1.09	$\pm$	0.15	&	(9.61	$\pm$	0.75)$\times 10^{47}$	&	4.20 	$\pm$	0.60 	&	6.53 	$\pm$	0.51 \\
200306C	&	...	&	1.58	&	2.80 	&	0.43	$\pm$	0.02	&	(1.44	$\pm$	0.10)$\times 10^{48}$	&	8.67 	$\pm$	0.54 	&	8.47 	$\pm$	0.35 \\
\hline
\end{tabular}
\note{A putative redshift $z=3.74$ based on five unambiguous observations is used to perform the cosmological correction for GRB\,060413, GRB\,071118, and GRB\,200306C.}
\note{\bf  Reference \rm (1) \cite{2005GCN..3709....1C,2005GCN..3710....1R,2005GCN..3716....1H,2005GCN..3732....1P,2005GCN..3746....1D}; (2) \cite{2006GCN..5237....1L}; (3) \cite{2011GCN.12648....1V}; (4) \cite{2014GCN.15924....1D,2014GCN.15936....1J}; (5) \cite{2020GCN.29100....1M}. \rm}
\end{table}

\begin{table}
\renewcommand{\thetable}{\arabic{table}}
\centering
\caption{The mass trace in each flare. }
\centering
\label{Tab:injection_mass}
\footnotesize
\begin{tabular}{lcccccccc}
%\tablewidth{0pt}
\hline\hline											
GRB-flares	&	$t_p$	&	$E_{\rm fla}$~($\rm ergs$)	&	$M_{\rm acc}~(M_\odot)$	&	$M_{\rm jet}~(M_\odot)$	&	$M_{\rm disk,loss}~ (M_\odot)$	\\
050730-$\rm\uppercase\expandafter{\romannumeral1}$	&	230	&	1.11$\times 10^{51}$	&	7.72$\times 10^{-3}$	&	9.28$\times 10^{-6}$	&	8.65$\times 10^{-3}$	\\
050730-$\rm\uppercase\expandafter{\romannumeral2}$	&	435	&	3.20$\times 10^{51}$	&	2.22$\times 10^{-2}$	&	2.67$\times 10^{-5}$	&	2.48$\times 10^{-2}$	\\
050730-$\rm\uppercase\expandafter{\romannumeral3}$	&	677	&	1.88$\times 10^{51}$	&	1.31$\times 10^{-2}$	&	1.57$\times 10^{-5}$	&	1.46$\times 10^{-2}$	\\
060607A-$\rm\uppercase\expandafter{\romannumeral1}$	&	95	&	1.05$\times 10^{51}$	&	7.30$\times 10^{-3}$	&	8.78$\times 10^{-6}$	&	8.18$\times 10^{-3}$	\\
060607A-$\rm\uppercase\expandafter{\romannumeral2}$	&	263	&	2.46$\times 10^{51}$	&	1.70$\times 10^{-2}$	&	2.05$\times 10^{-5}$	&	1.91$\times 10^{-2}$	\\
111209A	&	2027	&	2.06$\times 10^{51}$	&	1.43$\times 10^{-2}$	&	1.72$\times 10^{-5}$	&	1.60$\times 10^{-2}$	\\
140304A-$\rm\uppercase\expandafter{\romannumeral1}$	&	327	&	4.58$\times 10^{51}$	&	3.17$\times 10^{-2}$	&	3.81$\times 10^{-5}$	&	3.55$\times 10^{-2}$	\\
140304A-$\rm\uppercase\expandafter{\romannumeral2}$	&	820	&	2.70$\times 10^{51}$	&	1.87$\times 10^{-2}$	&	2.25$\times 10^{-5}$	&	2.10$\times 10^{-2}$	\\
201221A	&	4152	&	2.27$\times 10^{51}$	&	1.58$\times 10^{-2}$	&	1.89$\times 10^{-5}$	&	1.76$\times 10^{-2}$	\\
\hline											
060413	&	637	&	1.90$\times 10^{51}$	&	1.32$\times 10^{-2}$	&	1.58$\times 10^{-5}$	&	1.47$\times 10^{-2}$	\\
071118	&	593	&	1.63$\times 10^{51}$	&	1.13$\times 10^{-2}$	&	1.36$\times 10^{-5}$	&	1.26$\times 10^{-2}$	\\
200306C	&	1075	&	8.67$\times 10^{50}$	&	6.00$\times 10^{-3}$	&	7.22$\times 10^{-6}$	&	6.72$\times 10^{-3}$	\\
\hline																						
\end{tabular}
\note{}{}{The radius and the mass of the magnetar are set as $R=12~\rm km$ and $M=1.4~M_{\odot}$, respectively. The total mass loss of the disk is marked as $M_{\rm disk,loss}$.}
%\tablenotetext{}{The parameters of neutron star are set as $R=12~\rm km$, $M=2.37M_{\odot}$.} Furthermore, we consider that electromagnetic radiation and kinetic energy get same portion for a single flare.
\end{table}

\subsection{Observations VS Parameters}

\begin{figure}[h]
\centering
\includegraphics[angle=0,scale=0.2]{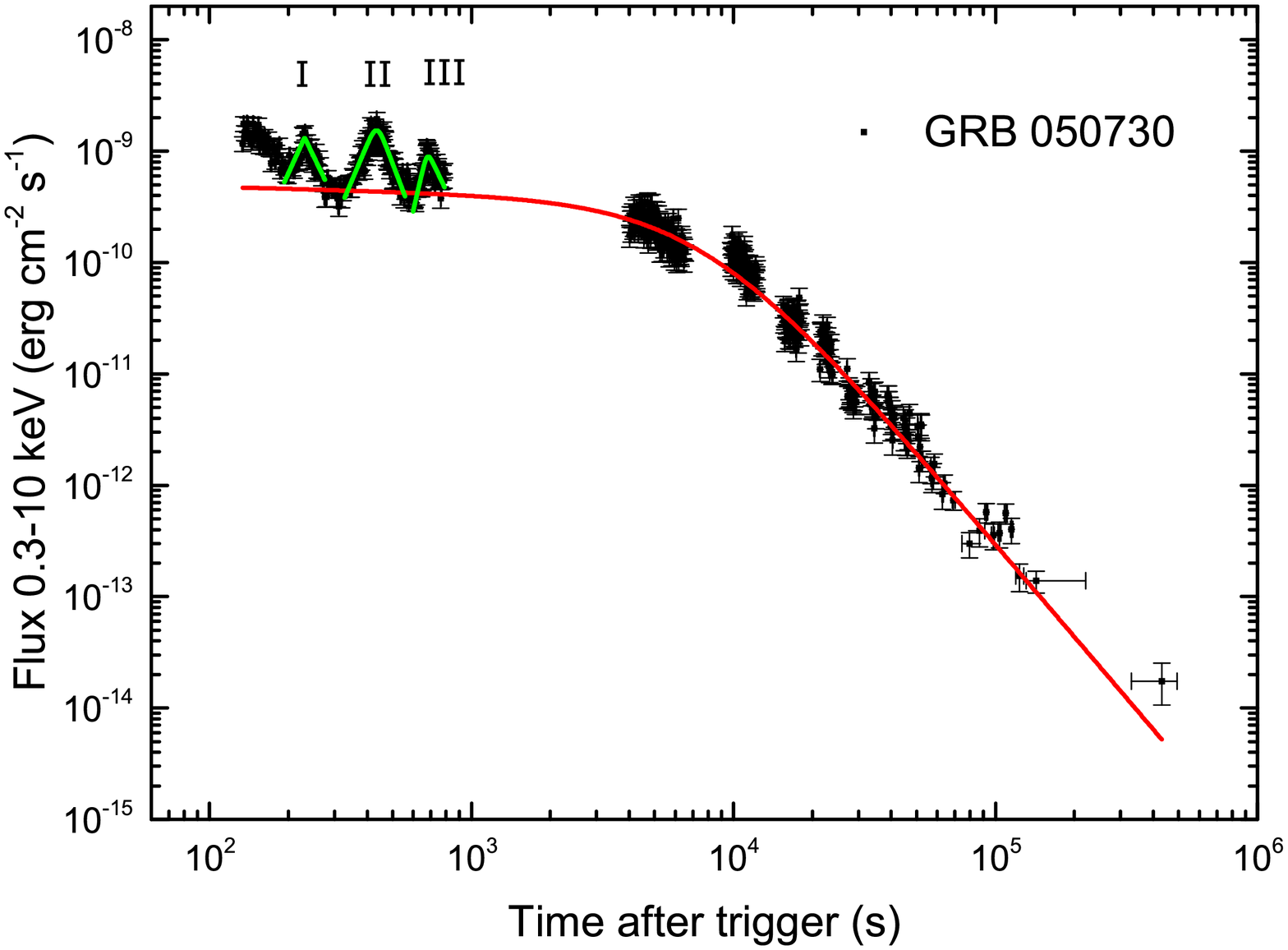}
\includegraphics[angle=0,scale=0.2]{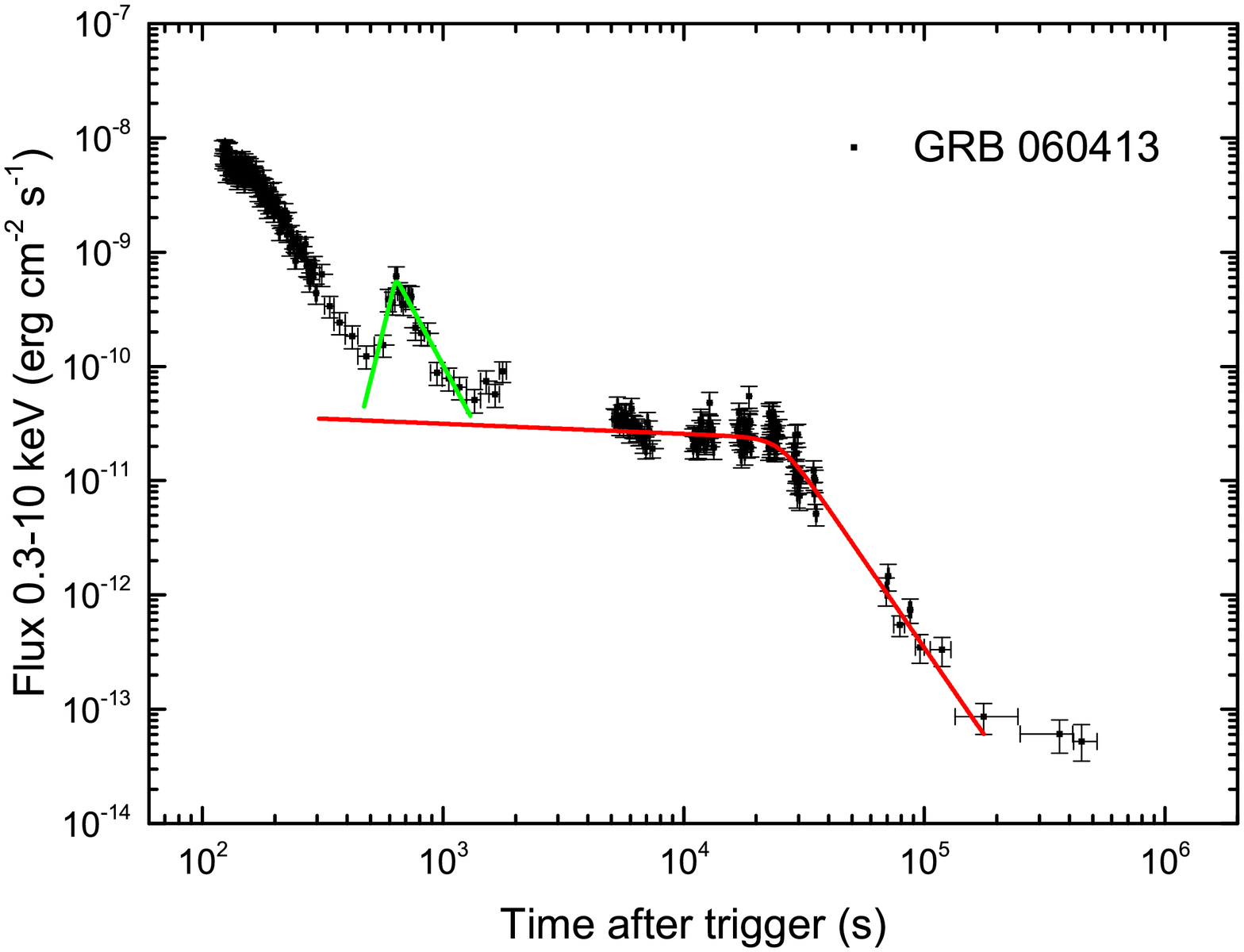}
\includegraphics[angle=0,scale=0.2]{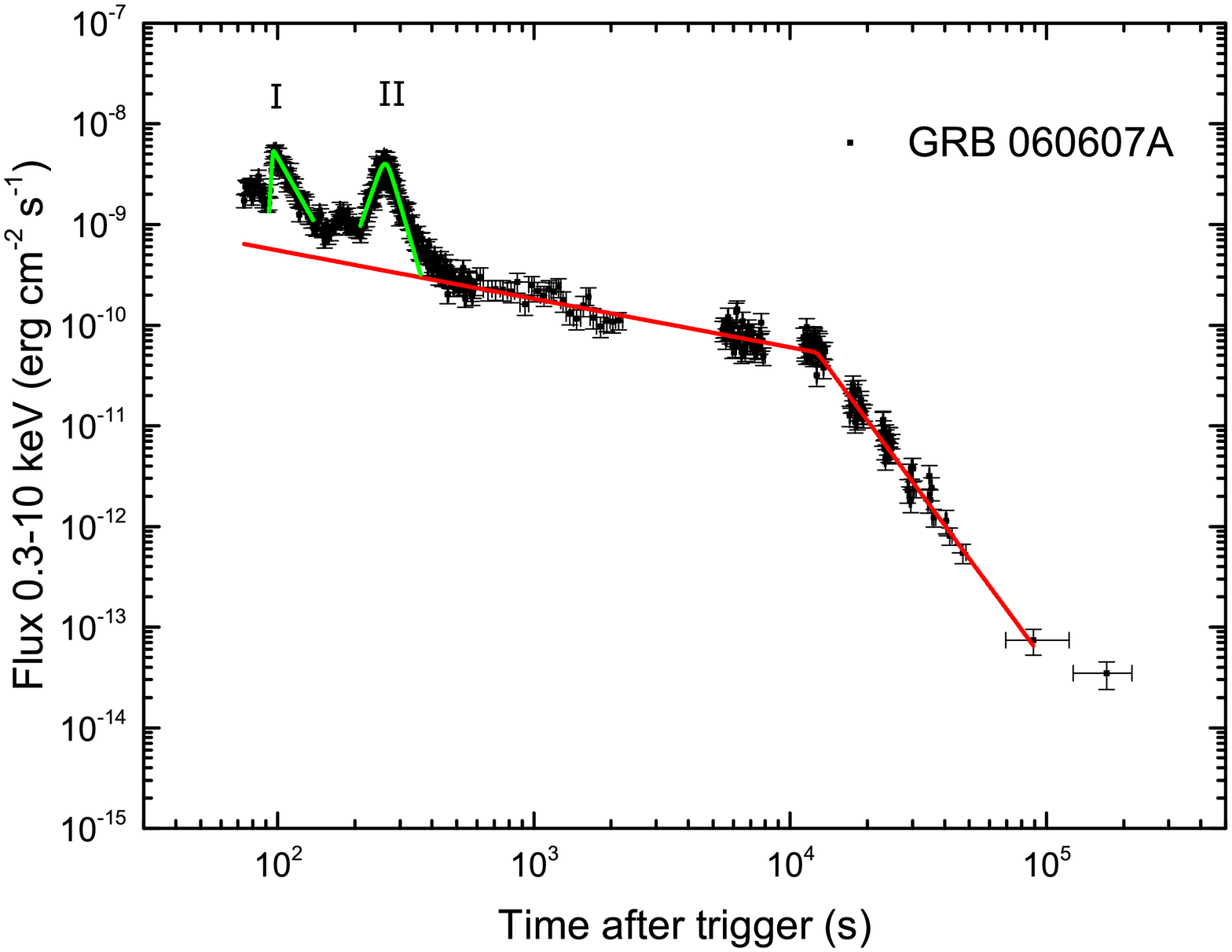}
\includegraphics[angle=0,scale=0.2]{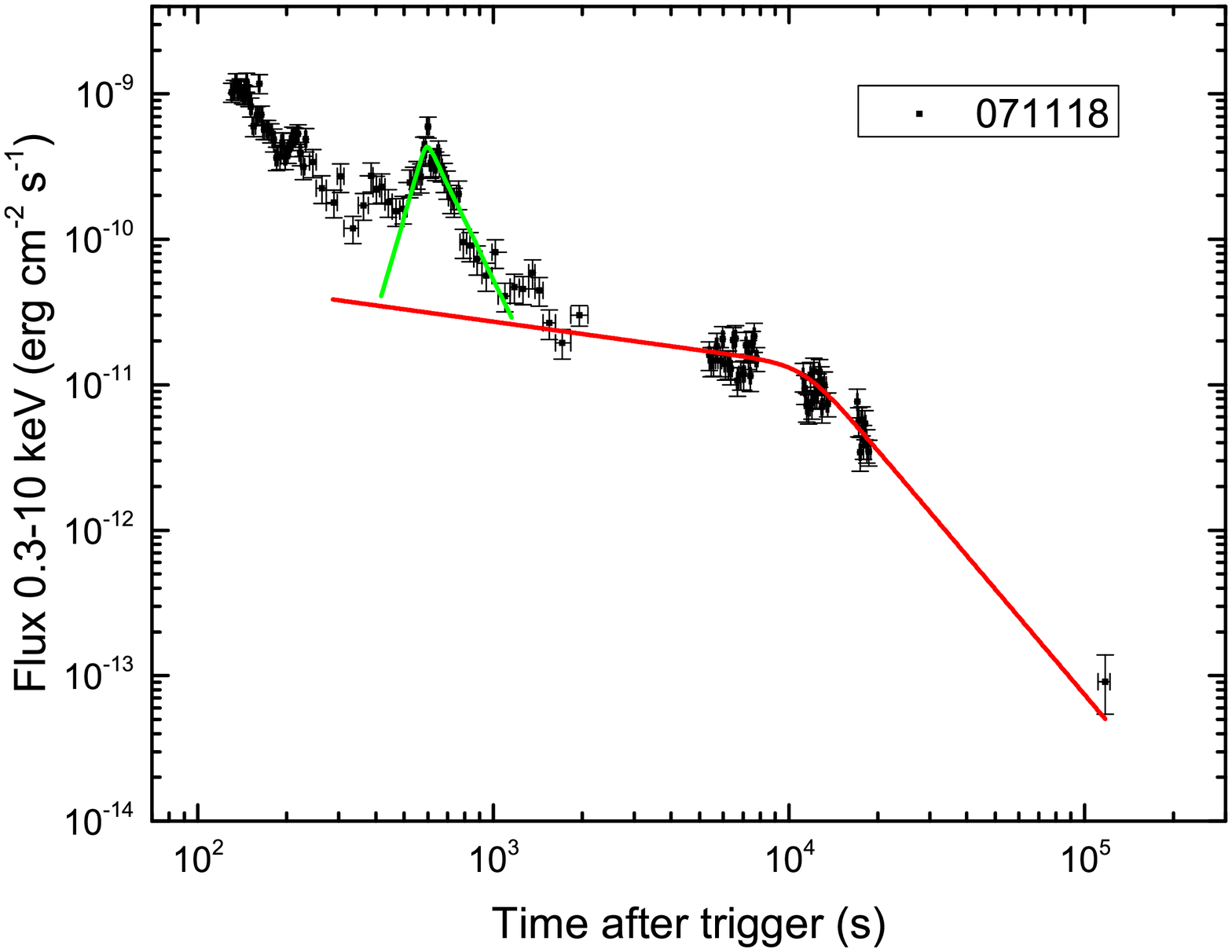}
\includegraphics[angle=0,scale=0.2]{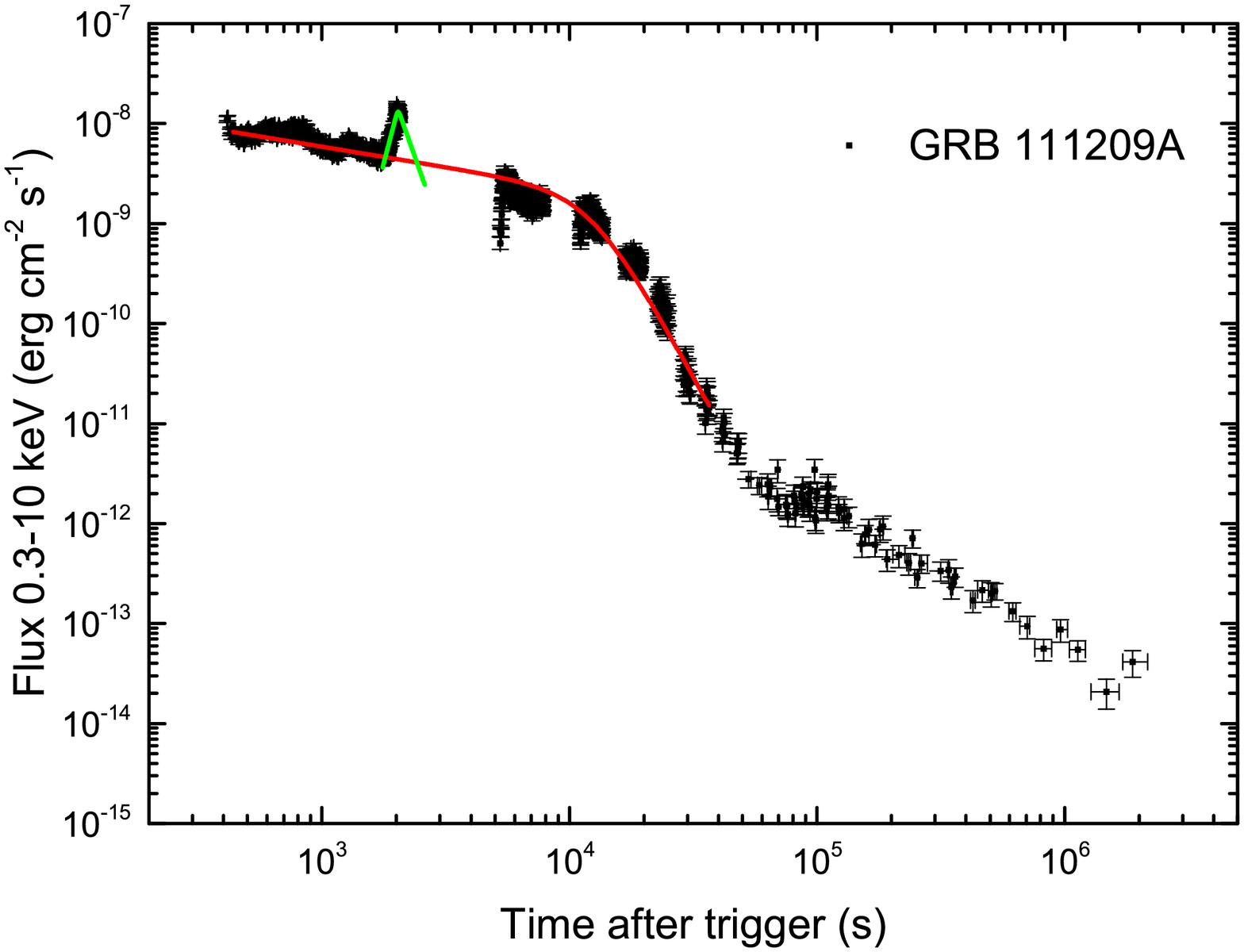}
\includegraphics[angle=0,scale=0.2]{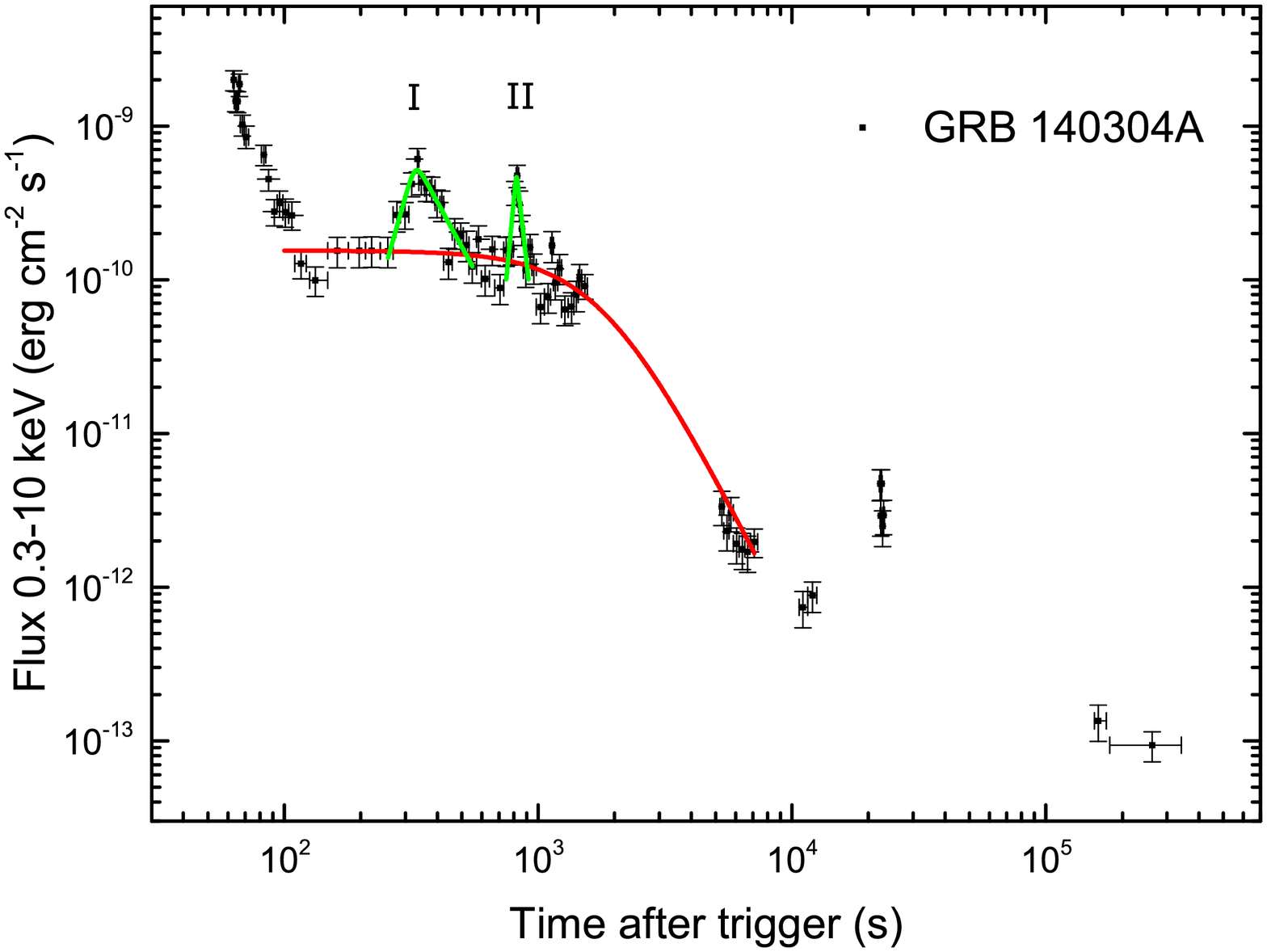}
\includegraphics[angle=0,scale=0.2]{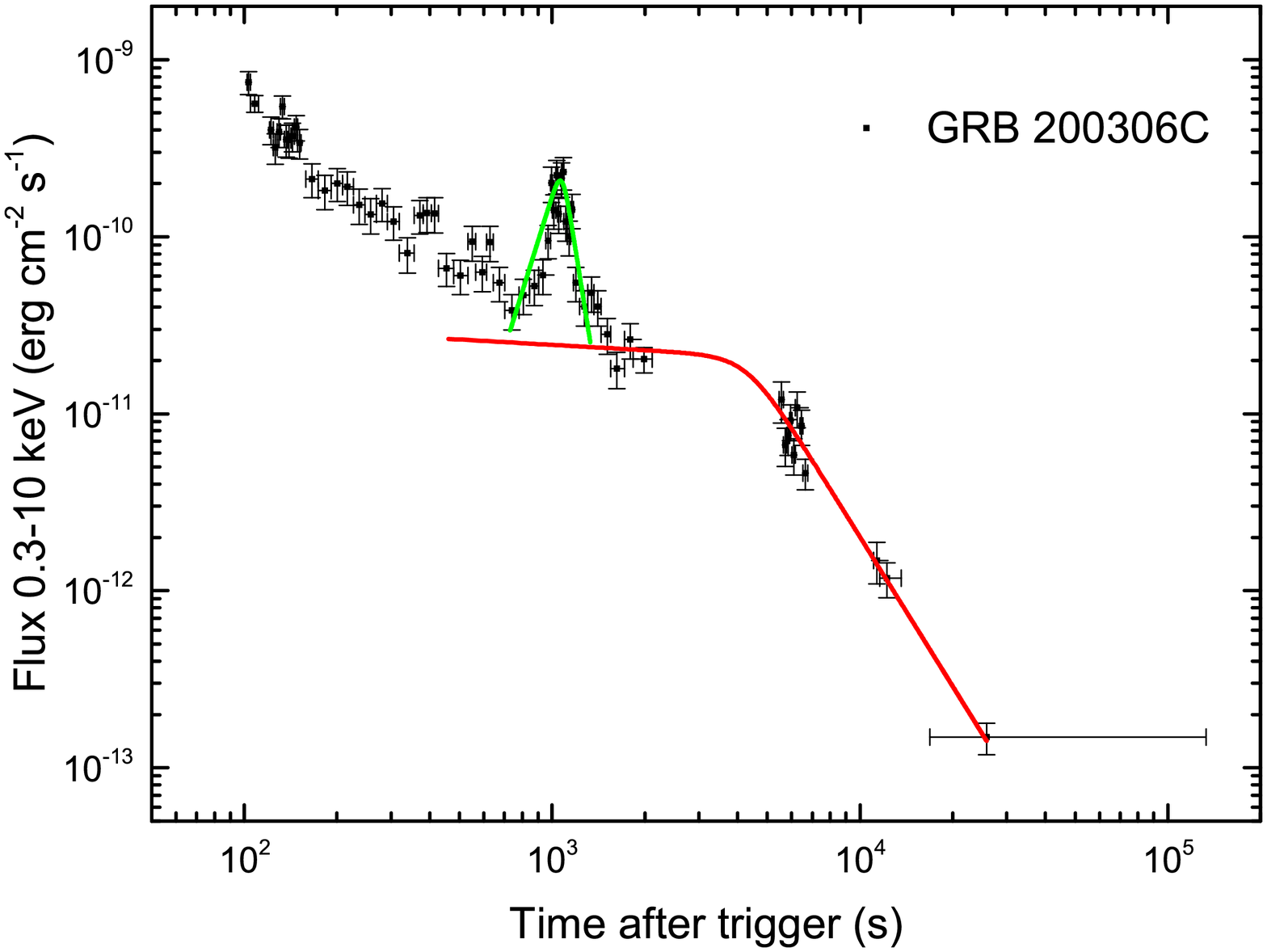}
\includegraphics[angle=0,scale=0.2]{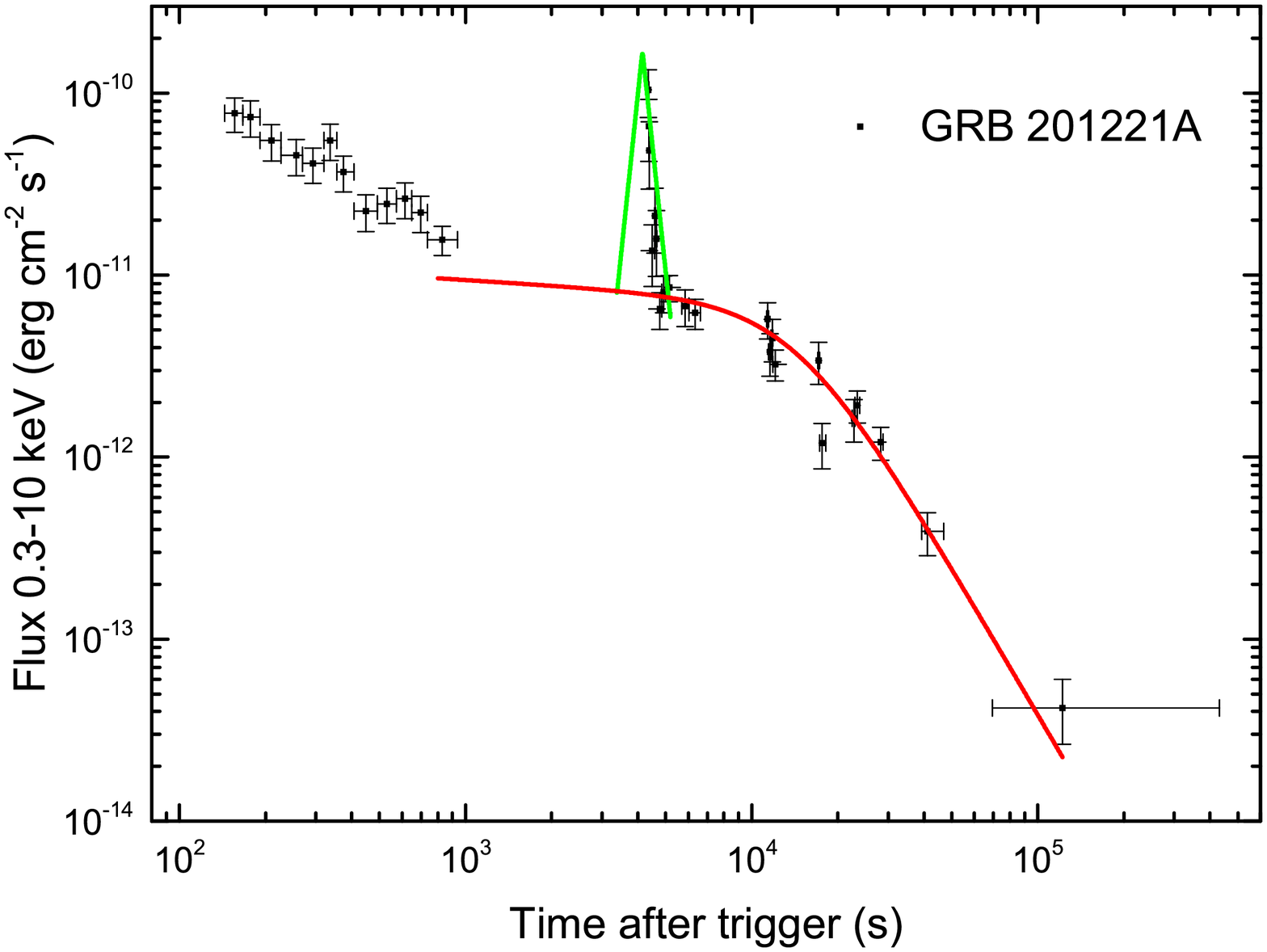}
\caption{The lightcurves of selected cases that X-ray flares raising upon a magnetar plateau. The magnetar plateaus are exhibited with a red line, and the prominent early flares are marked with green line, the putative profiles are used to give a demonstration for inadequate observations of GRB\,111209A and GRB\,201221A. Here, if the two or more flares are exhibited in one case, we donate the sign $\rm\uppercase\expandafter{\romannumeral1}$, $\rm\uppercase\expandafter{\romannumeral2}$, and $\rm\uppercase\expandafter{\romannumeral3}$ to distinguish these flares.}
\label{fig:samples}
\end{figure}

The selected cases present a prominent sharp decay ($\alpha_2<-2$) following a X-ray plateau ($\alpha_1>-0.75$). Ahead of the sharp decay, the remarkable X-ray flares raise upon the X-ray plateau. The lightcurves of the selected cases are shown in Figure \ref{fig:samples}, and the derived parameters of the MD radiation are organized in Table \ref{Tab:Magnetar}.
In these collected cases, five in eight have an unambiguous redshift measurement, four of it are large than 3.
The mean redshift of the five cases is $3.74$, it is higher than the mean redshift $z = 2.22$ for all of the redshift-measured GRBs that is detected by {\em Swift} prominently, this interesting phenomenon was also noticed by \cite{2010MNRAS.402..705L}. Here, the mean redshift $z=3.74$ is adopted to achieve the cosmological correction for the rest of the three GRBs.
The time average photon index $\Gamma$ of X-ray afterglow for each GRB is listed in column (3), and its median is 1.58. The isotropic X-ray characteristic luminosity of the magnetar spindown process is listed in column (6), and the characteristic spindown time scale in the observer frame is listed in column (5). The surface magnetic field and the initial spin period are presented in column (7) and (8), respectively.

\begin{figure}[h]
\centering
\includegraphics[angle=0,scale=0.35]{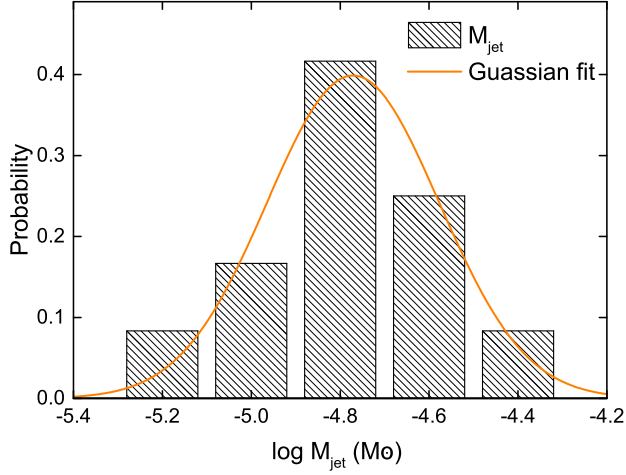}
\caption{The mass loading distribution of the observed 12 X-ray flares. A gaussian profile is used to show the distribution of mass loading and is exhibited with an orange line.}
\label{fig:Mjet}
\end{figure}

There are two or more flares identified in three GRBs (GRB\,050730, GRB\,060607A, and GRB\,140304A), we collect the peak time $t_p$ of flares and list them in Table \ref{Tab:injection_mass} and column (2). In each GRB, the time interval of the first flare to the trigger time almost equal to the time interval of the adjacent two X-ray flares quantitatively.
The total mass loading in each flare is listed in Table \ref{Tab:injection_mass}. Its probability distribution is also presented in Figure \ref{fig:Mjet}, and the centre of the gaussian profile is $10^{-4.77\pm0.19}~ M_{\odot}$.% In Table,  which is list $\dot{M}_{\rm dis,p}$.}

Following the time interval of the two adjacent flares, the average mass flow rates of the disk for three multi-flare GRBs are estimated to be $3.53\times 10^{-4}~M_\odot~\rm s^{-1}$, $4.23\times 10^{-4}~M_\odot~\rm s^{-1}$, and $4.33\times 10^{-4}~M_\odot~\rm s^{-1}$ for GRB\,050730, GRB\,060607A, and GRB\,140304A, respectively. Therefore, by using Eq. (\ref{eq:Alfven_radius}), the average sizes of the magnetosphere for corresponding GRBs are $5.01~\rm \times10^{6} cm$, $6.45~\rm \times10^{6} cm$, and $1.09~\rm \times10^{7} cm$, respectively.

\section{Conclusions and Discussions}
\label{sect:CaD}
%Long-duration plateau accompanying sharp decay in the X-ray afterglows is thought to be robust evidence of the spindown process of a newly born millisecond magnetar. X-ray flares discovered at the X-ray afterglow have been discussed for the activities of newly born millisecond magnetar (Dai et al. 2006).
%Some of these flares exhibit at the early time of the magnetar plateau, hinting the magnetar that the MD radiation is ongoing can not resist the mass flow onto the surface of it efficiently.
In this paper, we argue that the X-ray flares raising upon a magnetar plateau can be used to connect the accretion process of a magnetar and lead an implication for the existence of a disk surrounding a rapidly rotating and highly magnetized newborn NS.
In this scenario, the repeating accretion process is investigated in multi-flare GRB\,050730; the accreting induced variation of the magnetic field is discussed in GRB\,111209A, and hundreds of seconds re-magnetized process in an accreting magnetar scenario is implied.
In the selected cases, three of it show multiple flares in one GRB. In each GRB, the time interval of the first flare to trigger time almost equal to the time interval of the adjacent two X-ray flares quantitatively. In this scenario, by adopting magnetar mass $M=1.4~ M_\odot$ and radius $R=12~\rm km$, the average mass flow rates of the disk are $3.53\times 10^{-4}~M_\odot~\rm s^{-1}$, $4.23\times 10^{-4}~M_\odot~\rm s^{-1}$, and $4.33\times 10^{-4}~M_\odot~\rm s^{-1}$ for GRB\,050730, GRB\,060607A, and GRB\,140304A, respectively, and the corresponding average sizes of the magnetosphere of central magnetar are $5.01~\rm \times10^{6} cm$, $6.45~\rm \times10^{6} cm$, and $1.09~\rm \times10^{7} cm$.
A sample analysis that contains 8 GRBs within 12 flares shows that the total mass loading in single flare is $\sim 2\times 10^{-5}~M_{\odot}$.
In the lost mass of a disk, there are about 0.1\% portion used to feed the collimated jet.
%in which less than 0.2\% portion is used to feed the collimated jet.
The details are presented in Table \ref{Tab:injection_mass}.
Our mass loading result is compatible with the study in \cite{2009ApJ...707.1623M}. However, since the prominent high redshift ($z>3$ four in five) is given, the selective effect or more deep physical origination may be hinted.

Comparing to stellar fragment fallback determined accretion \citep[e.g.,][]{2021ApJ...914L...2L}, the MHD instability induced periodic accretion process have a shorter time interval for two successive accretion, for this scenario, the inner disk ram pressure can be released timely.
Furthermore, the observed periodic accretion process may imply the disk has reached a quasi-steady state (The disk may erratic when it born in the collapse of a massive star). Just as the simulation in \cite{2021ApJ...907...87L}, after a significant accretion, the fastness parameter keeps close to 1, the propeller and the accretion effect is comparable. In this scenario, the MHD instability leads the oscillation of disk and triggers the short-duration accretion.
Even the periodic accretion process has been studied for several decades \citep[e.g.,][]{1991ApJ...376..214B,1997ApJ...489..890M,2006ApJ...646..304U,2009MNRAS.399.1802R,2018NewA...62...94R}, a simple analytical formula for each step of the four-step accretion cycle and the jet launch mechanism is hard to be organized still. Hence, the time domain lightcurve is not explored in this work. In the future, by comprehensive consider both the magnetar-disk interaction, and the spin and magnetic field determined jet launch mechanism \citep[e.g., magnetic pressure powered collimated jet;][]{1995MNRAS.275..244L,1997ApJ...489..199G,1999ApJ...524..142G}, a time domain analytical solution would paint these quasi-steady magnetar-disk system vividly.

There are about $0.01~M_{\odot}$ materials needed in a single event, which can be a trouble for a supernova that has exploded more than hundreds or thousands of seconds \citep{2008Sci...321..376K}. The magnetosphere collects the fallback materials onto the surrounding disk would boost this process.
As a reward, the accompanying disk helps the magnetar to store the rotational energy and returns it when the spin decreases because of an MD radiation or gravitational radiation. This scenario can be used to understand why a high efficiency of converting the rotational energy to the observed X-ray emission was found in \cite{2014ApJ...785...74L}, by comparing with a modified efficiency format, e.g., $\eta = \int{L_{X}}dt/(E_{rot,NS}+E_{disk})$, where $L_{X}$, $E_{rot,NS}$, $E_{disk}$ are the MD luminosity in X-ray band, the rotational energy of NS, and the energy stored in the surrounding disk, respectively (Zheng et al. 2022, in preparation).
Given the two-component outflow scenario, the re-brightening following a sharp decay, e.g., GRB\,111209A, can be explained as the catching up of delay conical wind. Considering the evolution of central magnetar, that a small block mass fallback accretion due to the magnetosphere gets a significantly shrinking may also give a reasonable solution for later re-brightening. Therefore, for a later re-brightening following magnetar plateau scenario, at least in some cases, the final stage of the GRB central engine can be an NS rather than a BH \citep{2020ApJ...895...46L}, further deep observations are expected to reveal the mask of the central engine of GRBs.

\begin{acknowledgements}
We acknowledge the use of the public data from the {\em Swift} data archive and and the UK Swift Science Data Center. We thank the anonymous referee for helpful recommendations to enhance this work, and the selfless discussions with professor Hou-Jun L\"{u} and En-Wei Liang. This work is supported by the National Natural Science Foundation of China (Grant No. U1938201), the Guangxi Science Foundation the One-Hundred-Talents Program of Guangxi colleges, and Innovation Project of Guangxi Graduate Education (Grant No. YCBZ2020025).

\end{acknowledgements}

\label{lastpage}

\end{document}